\documentclass[12pt,preprint]{aastex}

\shorttitle{Granular scale flux cancellation}
\shortauthors{Kubo et al.}

\begin{document}

\title{Granular Scale Magnetic Flux Cancellations in the Photosphere}

\author{M. Kubo, B. C. Low, B. W. Lites}
\affil{High Altitude Observatory, National Center for
Atmospheric Research\altaffilmark{1}, P.O. Box 3000, Boulder, CO 80307, USA.}

\altaffiltext{1}{The National Center for Atmospheric Research is sponsored by the National Science Foundation}

\begin{abstract}
We investigate the evolution of 5 granular-scale magnetic flux
 cancellations just outside the moat region of a sunspot by using
 accurate spectropolarimetric measurements and G-band images with the
 Solar Optical Telescope aboard \textit{Hinode}. 
The opposite polarity magnetic elements approach a junction of the
 intergranular lanes and then they collide with each other there. 
The intergranular junction has strong red shifts, darker intensities
 than the regular intergranular lanes, and surface converging flows. 
This clearly confirms that the converging and downward convective
 motions are essential for the approaching process of the
 opposite-polarity magnetic elements.
However, motion of the approaching magnetic elements does not always match
 with their surrounding surface flow patterns in our observations.
This suggests that, in addition to the surface flows, subsurface 
 downward convective motions and subsurface magnetic connectivities are
 important for understanding the approach and collision of the opposite
 polarity elements observed in the photosphere.    
We find that the horizontal magnetic field appears between the
 canceling opposite polarity elements in only one event.
The horizontal fields are observed along the intergranular lanes with
 Doppler red shifts.
This cancellation is most probably a result of the submergence
 (retraction) of low-lying photospheric magnetic flux.
In the other 4 events, the horizontal field is not observed between the
 opposite polarity elements at any time when they approach and cancel
 each other. 
These approaching magnetic elements are more concentrated rather than
 gradually diffused, and they have nearly vertical fields even while
 they are in contact each other. 
We thus infer that the actual flux cancellation is highly time dependent
 events at scales less than a pixel of \textit{Hinode} SOT (about 200
 km) near the solar surface.   
\end{abstract}

\keywords{Sun: granulation --- Sun: magnetic fields --- Sun: photosphere}

\section{INTRODUCTION}
The magnetic fields in the photosphere are highly heterogeneous, as is
well known \citep[e.g.][]{Parker1979, Zwaan1987, Stenflo1989}.
These fields exist in a dynamical state in a spectrum of sizes, ranging
from the kilo-gauss fibrils with diameters of a few hundred kilo-meters,
through the mesoscale pores, to sunspots of various sizes.
In the turbulent convection zone immediately below the quiet-Sun
photosphere, the magnetic field is dominated by the fluid. 
Therefore, it is expected that the field distribution is organized into
the convective cells. 
This field distribution is necessarily complex with flux connectivities
of various scales, across individual and multiple cells.
Magnetic concentrations that have vertical and strong (kilo-gauss) field
are observed along the boundaries of the convective cells.
A most prominent feature in the quiet Sun is the network magnetic field
that partially outlines supergranular cells.
It is believed that the network magnetic field is formed by the
advection of internetwork fields via the supergranular flows.
Magnetic elements in internetwork areas tend to move toward the nearest network concentration
at a speed of about 0.2 km s$^{-1}$ \citep{deWijn2008}, and their rms
velocity is about 1.5 km s$^{-1}$ in internetwork areas \citep{Nisenson2003, deWijn2008}.
Velocities of 0.2 km s$^{-1}$ and 1.5 km s$^{-1}$ are similar to a
typical speed of supergranular flows and granular flows, respectively.

Down at the scales of a few hundred kilo-meters, equal to a few
photospheric density scale heights, magnetic elements of both polarities
form to often merge and collide with each other in the photospheric
fluid.
The net magnetic flux increases or decreases, depending on whether
like-polarity or opposite-polarity magnetic elements have merged.
In this paper, we report an interesting set of ``magnetic flux
cancellation'' events of this kind observed at the high spatial
resolution of the Solar Optical Telescope \citep[SOT;][]{Tsuneta2008}
aboard \textit{Hinode} \citep{Kosugi2007}.
The magnetic flux cancellation is a descriptive term to indicate a
mutual flux loss due to the apparent collision of the opposite-polarity
magnetic elements \citep{Martin1985}. 
This flux cancellation is essential to the process of replacement of old
magnetic flux with newly emerging flux in the quiet Sun on a timescale
of a few days \citep{Schrijver1997, Hagenaar2001}, and also to the
process of removal of sunspot magnetic flux from the photosphere
\citep{Kubo2008}. 

Various possible processes have been proposed to explain the observed
flux cancellation \citep[e.g.][]{Zwaan1987, Ryutova2003}, involving
submergence (retract) of $\Omega$-shaped loops or emergence of U-shaped
loops across the photosphere. 
In both cases, the canceling opposite-polarity magnetic elements
correspond to the two intersections of such loops with the photospheric layer.
The opposite-polarity magnetic elements disappear when the top of a
submerging $\Omega$-loop has passed through the photospheric layer
\citep[see Fig.2 in][]{Zwaan1987}. 
Alternatively, these elements disappear when the bottom of a rising
U-loop has passed clear through the photospheric layer.
When magnetic field lines have emerged into the chromosphere and corona,
they can hardly submerge back below the photosphere because of magnetic
buoyancy. 
Magnetic reconnection taking place within several scale
heights above the solar surface is probably needed to create low-lying
$\Omega$-loops whose magnetic tension force can then overcome the
magnetic buoyancy force \citep{Parker1975}.
In the photospheric magnetic reconnection cases, reconnection should be
most efficient around the temperature minimum region: about 600 km above
the solar surface \citep{Litvinenko1999, Takeuchi2001}. 
In contrast, both magnetic tension and buoyancy forces are directed
upward in the case of a U-loop rising through the photosphere.
Magnetic reconnection is therefore not crucial for the emerging
U-loops. 

An important observable signature for understanding flux
cancellation is the motion of the horizontal magnetic field connecting
the canceling magnetic elements. 
Horizontal magnetic fields have been observed between the
opposite-polarity magnetic elements during the cancellations of moving
magnetic features around a sunspot \citep{Chae2004}.
Similar horizontal fields have also been observed in events of
cancellations of pores and sunspots \citep{Kubo2007}.   
A flux cancellation without increase of the horizontal
field has also been reported for the moving magnetic features
\citep{Bellot2005}.  
Knowledge of the full vector field permits one to determine whether the
field geometry has $\Omega$-loop topology or U-loop topology, but in the
case of cancellation of small, isolated flux elements, such a
determination is often compromised by usual difficulty of resolving
the 180$\degr$ azimuth ambiguity.
Regarding the motions at such a cancellation site, \citet{Harvey1999}
show that the magnetic flux disappears in the chromosphere before it
does in the photosphere for at least about half of the cancellation
events.  
They suggest that magnetic flux is submerging in most, if not all, of
the cancellation sites.
On the other hand, both Doppler red shift \citep{Chae2004} and Doppler
blue shift \citep{Yurchyshyn2001} are reported in the cancellation sites.
The center-to-limb variations of the Doppler velocities at the polarity
inversion lines in the cancellation sites suggest that the observed
velocities with the spatial resolution of about 1$\arcsec$ mainly show 
the gas flows along the horizontal fields rather than actual emerging or
submerging motions of the field lines \citep{Kubo2007}.  
Observationally, the physical process of the magnetic flux cancellation
is still not well understood.

Recent observations with the high spatial resolution by \textit{Hinode}
SOT show that downward motions (red-shifts) are continuously observed during
the flux cancellation process \citep{Iida2010}. 
From the Stokes-V signals far from the disk center, they also suggest
that the canceling opposite polarity elements tend to have an
$\Omega$-shaped configuration rather than a U-shaped configuration.
In this paper, we investigate the detailed evolution of canceling
magnetic elements and their surrounding convective motions near the disk
center by using vector magnetic fields and velocity fields observed with
the SOT. 
We are particularly interested in the evolution that precedes the
cancellation in order to investigate why the opposite-polarity magnetic
elements approach and collide with each other.
For a full understanding of the phenomenon, this dynamical development
is as important as the flux removal process at the cancellation site.

\section{OBSERVATIONS AND ANALYSIS}
\begin{figure}
\epsscale{0.40}
\plotone{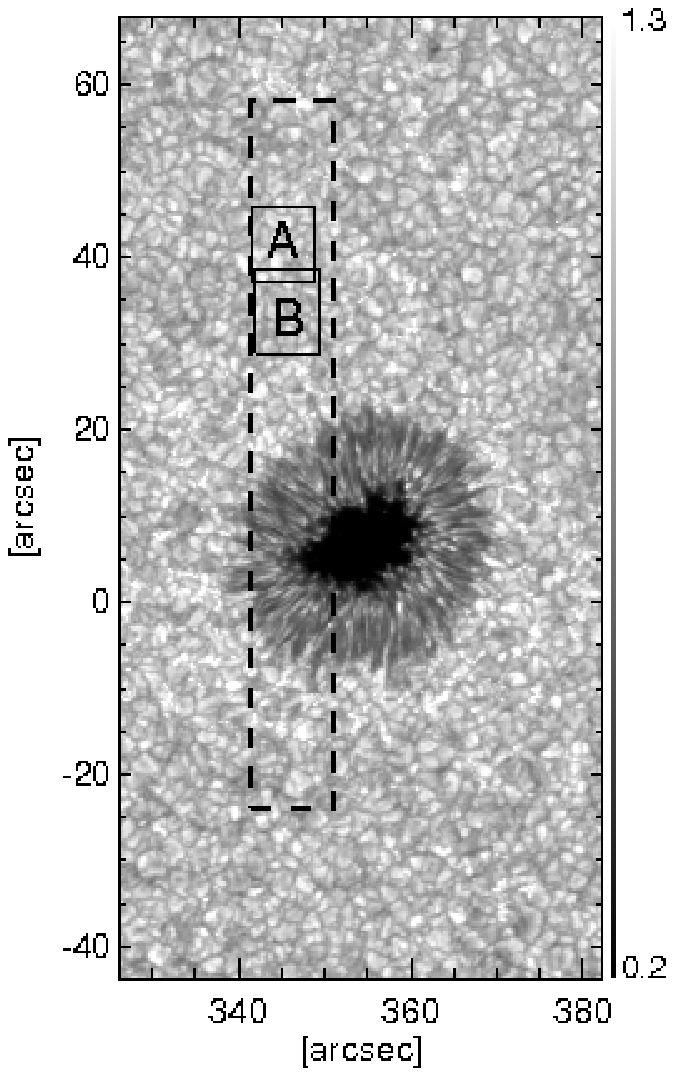}
\caption{
G-band image taken at 08:00:32 on 2007 March 2.
The G-band image is normalized to the mean intensity of the quiet area
 outside the sunspot.
The dashed box shows a scan area by the \textit{Hinode} SP.
The solid boxes labeled ``A'' and ``B'' are identical to the fields of
 view of Figs.~\ref{fig_lfov1} and \ref{fig_fov2_sp}, respectively.
}
\label{fig_full_fov}
\end{figure}

Our target flux-cancellation events occurred just outside an active
region NOAA 10944, as shown by the solid boxes in Figure~\ref{fig_full_fov}. 
We selected a data set simultaneously taken by the SOT spectropolarimeter
(SP) and filtergraph (FG) in their full spatial resolution modes.
The SP repeatedly scanned the same area with a field of view of
$10\arcsec \times 82\arcsec$ (the dotted box in Fig.~\ref{fig_full_fov})
for 4.5 hr from 06:35:32 on 2007 March 2. 
The SP measured Stokes I, Q, U, and V profiles across Fe \small{I} 630.1
nm and 630.2 nm lines.
A width of a slit was $0\arcsec.15$ and a pixel sampling along the
slit was $0\arcsec.16$. 
One scan took 5.5 minutes with an integration time of 4.8 s at
each slit position.
We estimated total circular polarization ($C_{tot}$)
and total linear polarization ($L_{tot}$) as follows:
\begin{equation}
C_{tot}= \frac{\int_{\lambda_0-21.6\thinspace\textrm{\scriptsize{pm}}}^{\lambda_0-4.32\thinspace\textrm{\scriptsize{pm}}}V(\lambda) \,d\lambda}{I_c\int_{\lambda_0-21.6\thinspace\textrm{\scriptsize{pm}}}^{\lambda_0-4.32\thinspace\textrm{\scriptsize{pm}}}\,d\lambda},
\end{equation}
\begin{equation}
L_{tot}= \frac{\int_{\lambda_0-21.6\thinspace\textrm{\scriptsize{pm}}}^{\lambda_0+21.6\thinspace\textrm{\scriptsize{pm}}}\sqrt{Q^2(\lambda)+U^2(\lambda)} \,d\lambda}{I_c\int_{\lambda_0-21.6\thinspace\textrm{\scriptsize{pm}}}^{\lambda_0+21.6\thinspace\textrm{\scriptsize{pm}}}\,d\lambda},\\
\end{equation}
where $\lambda_0$ is the center of Fe \small{I} 630.2 nm line in each pixel,
and $I_c$ is the local continuum intensity.
The continuum intensity is averaged over the Stokes I profile from
$\lambda_0+43.2$ pm to $\lambda_0+64.8$ pm. 
In the weak field approximations \citep{Jefferies1989}, $C_{tot}$ is
proportional to the longitudinal magnetic flux density ($fB_{L}$) and
$L_{tot}$ has the relation to the transverse flux density as $L_{tot}
\sim fB_T^2$, where $B_L$, $B_T$, $f$ are the longitudinal field
strength, transverse field strength, and filling factor, respectively.   
The observed area is located not far from the disk center (S06W21).
We can consider the longitudinal and transverse fields to be the vertical
and horizontal fields with respect to the solar surface, respectively.
In addition to the magnetic field, we compute the Doppler velocity
with respect to the average of Doppler velocities in the quiet area. 

During the repeated scans by the SP, the FG took a time
series of G-band images with a pixel scale of 0$\arcsec$.05448.
The cadence is 1 minute, and the field of view is
$56\arcsec\times112\arcsec$. 
After applying a sub-sonic filter for the time series of G-band images,
we estimated granular flow patterns with the local correlation tracking
technique \citep[LCT;][]{November1988}. 
The apodization window of the LCT had a Gaussian shape with the FWHM of
32 pixels (1$\arcsec$.7).
The LCT flow maps are calculated from the series of G-band images.
The cadence of the LCT flow maps is therefore 1 minute.
Drift of the image, arising from the correlation tracker
\citep{Shimizu2008}, was removed by aligning the sunspot centers in the 
G-band images at different times.
After removing the image drift, both G-band images and flow maps were
averaged over the period corresponding to the duration of an SP map
(5.5 minutes).
The SP maps were aligned to the averaged G-band images at the time
closest to the midpoint of the SP maps.
The alignment was performed by the image cross-correlation between the
averaged G-band images and the SP continuum intensity maps.
Thus, the image drift in the time series of the SP maps was removed as
well as the G-band images.

\section{RESULTS}
Here we present the evolution of flux cancellation events in Region A
and Region B of Figure~\ref{fig_full_fov}.
We shall henceforth use simplified terminology of calling the
positive-polarity and negative-polarity magnetic elements simply as the
positive and negative elements, respectively.

\subsection{Region A}
\begin{figure}
\epsscale{1.00}
\plotone{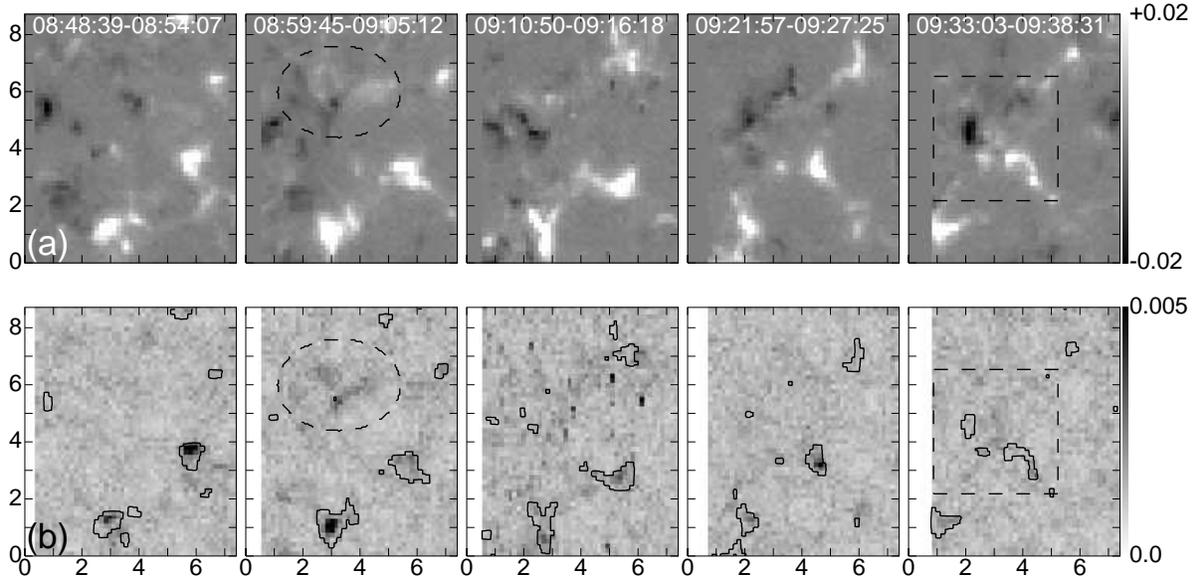}
\caption{
Time series of (a) the total circular polarization ($C_{tot}$) and (b) the
 total linear polarization ($L_{tot}$) before the flux cancellation in
 Region A of Figure~\ref{fig_full_fov}. 
White (black) indicates positive (negative) polarity in panel \textit{a}.
The contours in panel \textit{b} represent the $\pm 0.01 C_{tot}$ levels.
The units of vertical and horizontal axes are in arcseconds.
The dashed circle indicates a flux emerging site.
The dashed box in the final frame includes our target flux cancellation
 event, and is identical to the field of view of Fig.~\ref{fig_sfov1}.
}
\label{fig_lfov1}
\end{figure}

The first event is an elemental flux cancellation in Region A. 
Figure~\ref{fig_lfov1} shows the developments of this region prior to
the start of the cancellation event.
Fuzzy positive (white) and negative (black) elements appear within the
dashed circle in the map of the total circular polarization ($C_{tot}$).
In this circle, an increase of the total linear polarization ($L_{tot}$)
is also observed.   
These results indicate that the opposite-polarity elements emerge into
the photosphere as a pair whereas the canceling magnetic elements in the
dashed box do not originally emerge as a pair.
The emerged negative elements merge together to consolidate
into a stronger negative element that subsequently approach and
mutually cancel with a positive element.
Even when the negative element is close to the
positive element, there is only a slight increase of $L_{tot}$
in a part of the region between them.

\begin{figure}
\epsscale{1.00}
\plotone{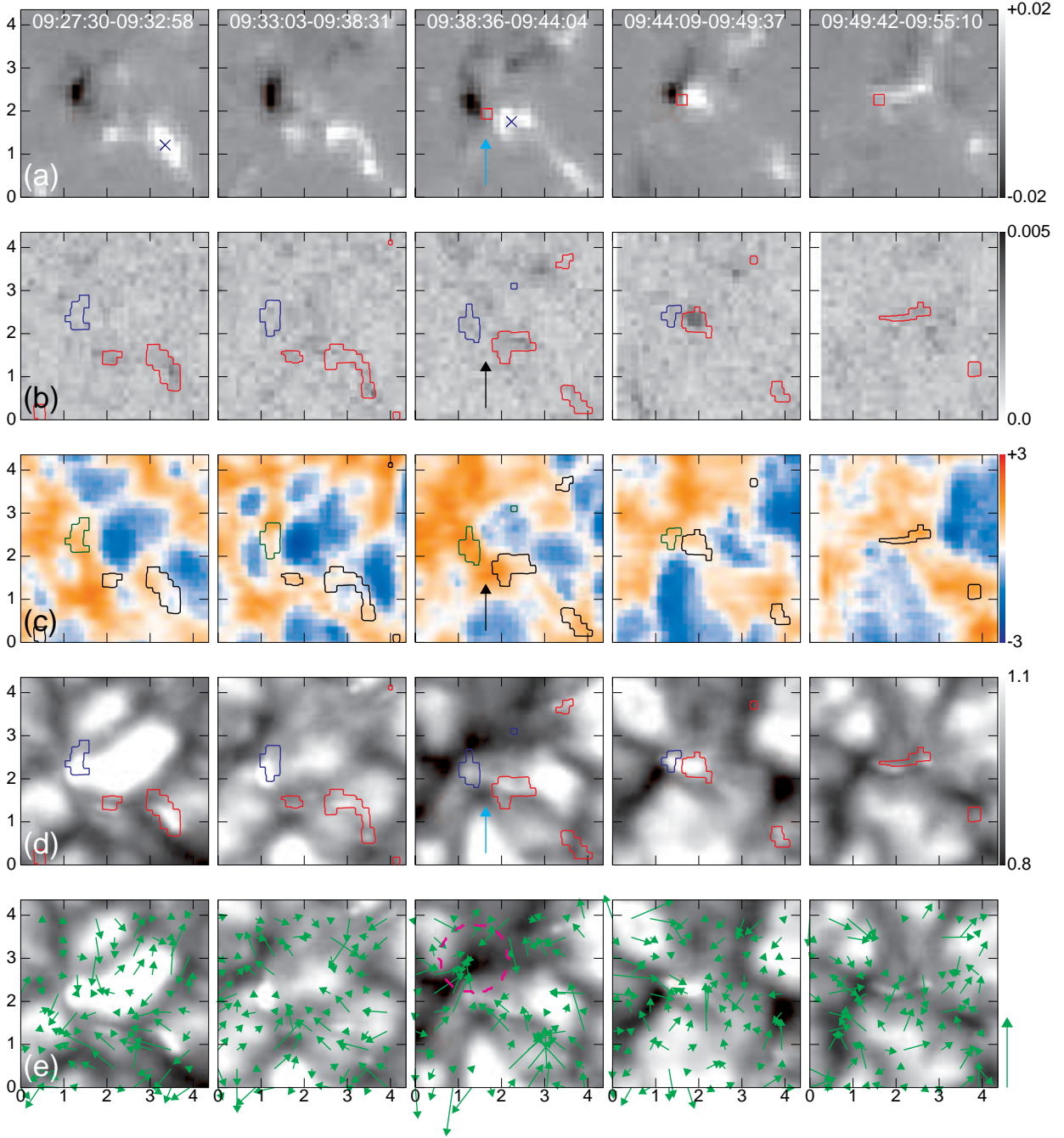}
\caption{
(a) Total circular polarization ($C_{tot}$), (b) total linear
 polarization ($L_{tot}$), (c) Doppler velocity in units of km s$^{-1}$,
 (d) normalized G-band intensity, and (e) flow map during the flux
 cancellation in Region A.    
The upward arrows in panels \textit{a-d} indicate the strong
 red-shifted area, and the contours are $\pm0.01$ of $C_{tot}$.
The cross symbols in panel \textit{a} represent centers of the
 positive-polarity magnetic element to be canceled.
The arrows in panel \textit{e} show the horizontal velocities
 derived with the local correlation tracking method.
The upward arrow on the right-hand side of the panel is 1.5 km s$^{-1}$.
The background of panel \textit{e} is identical to that of panel \textit{d},
 and the dashed circle surrounds the area with strong converging flows. 
}
\label{fig_sfov1}
\end{figure}

\begin{figure}
\epsscale{1.00}
\plotone{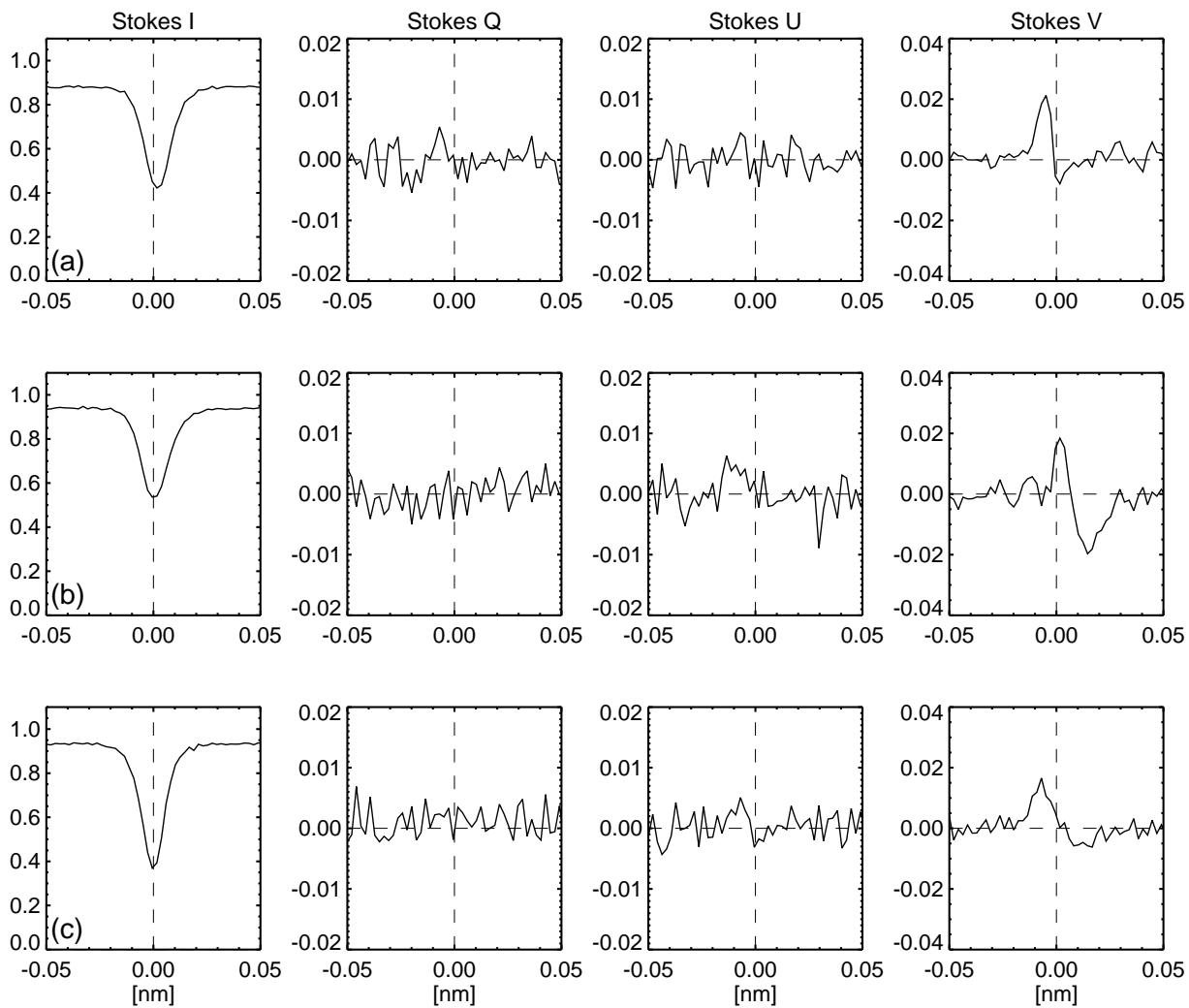}
\caption{
Stokes profiles of the Fe \small{I} 630.2 nm line at the pixels
 represented by squares in panel \textit{a} of Fig.~\ref{fig_sfov1}.  
The profiles in panels \textit{a-c} are taken at 09:39:39, 09:45:02, and
 09:50:30 on 2009 March 2, respectively. 
These profiles are normalized by the continuum intensity averaged over
 the quiet area. 
The vertical dashed line represents the averaged position of the line
 centers over the map. 
}
\label{fig_profile1}
\end{figure}

Figure~\ref{fig_sfov1}\textit{a} shows that, like the negative canceling
element, smaller positive elements around the cancellation site first
converge to form into a prominent positive element that then moves
toward the negative element.
We trace the center of the positive element.
The center (the cross symbol in Fig. 3\textit{a}) is defined as the
average of the position ($\bold{r}$) weighted by $C_{tot}$:
$\sum{C^i_{tot}}\bold{r}_i/\sum{C^i_{tot}}$.  
This center moves at about 1.4 km s$^{-1}$ during 11 minutes.
Note that the motion of 1.4 km s$^{-1}$ is similar to the rms velocity
of magnetic elements in internetwork areas \citep{Nisenson2003, deWijn2008}. 
The positive element collides with the negative magnetic
element at a time between the third and fourth frames. 
The negative element disappears from the photosphere by 10 minutes
after the start of the cancellation, whereas the positive element does
not completely disappear in this event. 
Even while the net flux of both negative and positive elements
is decreasing, $C_{tot}$ of each pixel in the colliding magnetic
elements does not decrease. 
Furthermore, both $C_{tot}$ and $L_{tot}$ slightly increase at the center
of the positive element at the fourth frame in Figure~\ref{fig_sfov1}. 
This increase arises from either the increase of the filling factor or the
field strength within the positive element.  
These results suggest that the colliding magnetic elements have nearly vertical
magnetic fields until these magnetic elements disappear form the
photosphere. 
On the other hand, no increase of $L_{tot}$  is observed between the
canceling opposite polarity elements during the whole flux cancellation
process. 
This means that the horizontal magnetic fields, which are expected in
both the submerging $\Omega$-loop model and the emerging U-loop model, do not
appear at the cancellation site.
Figure~\ref{fig_profile1} confirms that neither Stokes Q nor U signal
increases during the flux cancellation.  

The cancellation site is located at a junction of the intergranular
lanes (the upward arrow in Fig.~\ref{fig_sfov1}\textit{d}).
The intergranular junction is darker in G-band intensity than its nearby
intergranular lanes.   
Intergranular lanes have generally red shifts, and a larger red shift
with about 1.5 km s$^{-1}$ is observed at the intergranular junction 
just before the cancellation (Fig.~\ref{fig_sfov1}\textit{c}).
Interestingly, such a larger red shift is not observed there during the
cancellation.
The left panels of Figure~\ref{fig_profile1} also show that the red
shift of Stokes I profile is only seen in the top panel taken just
before the cancellation.     
This means that the observed large red shift at the cancellation site
does not result from this flux cancellation process. 
Instead, we may infer that the opposite polarity elements approach the
intergranular junction with the large red shift. 
The intergranular junction tends to have a converging flow rather than a
diverging flow, but the flow pattern is not clear in this event.
The size of the intergranular junction is too small to examine the flows
derived by the LCT method because the LCT window is 1$\arcsec$.7. 
One interesting result is that the negative element moves toward
such a small, less clear converging area although a large, strong
converging area with the darker intensity is also located near it, as
shown by the dashed circle in Figure~\ref{fig_sfov1}\textit{e}.

\subsection{Region B}
\begin{figure}
\epsscale{1.00}
\plotone{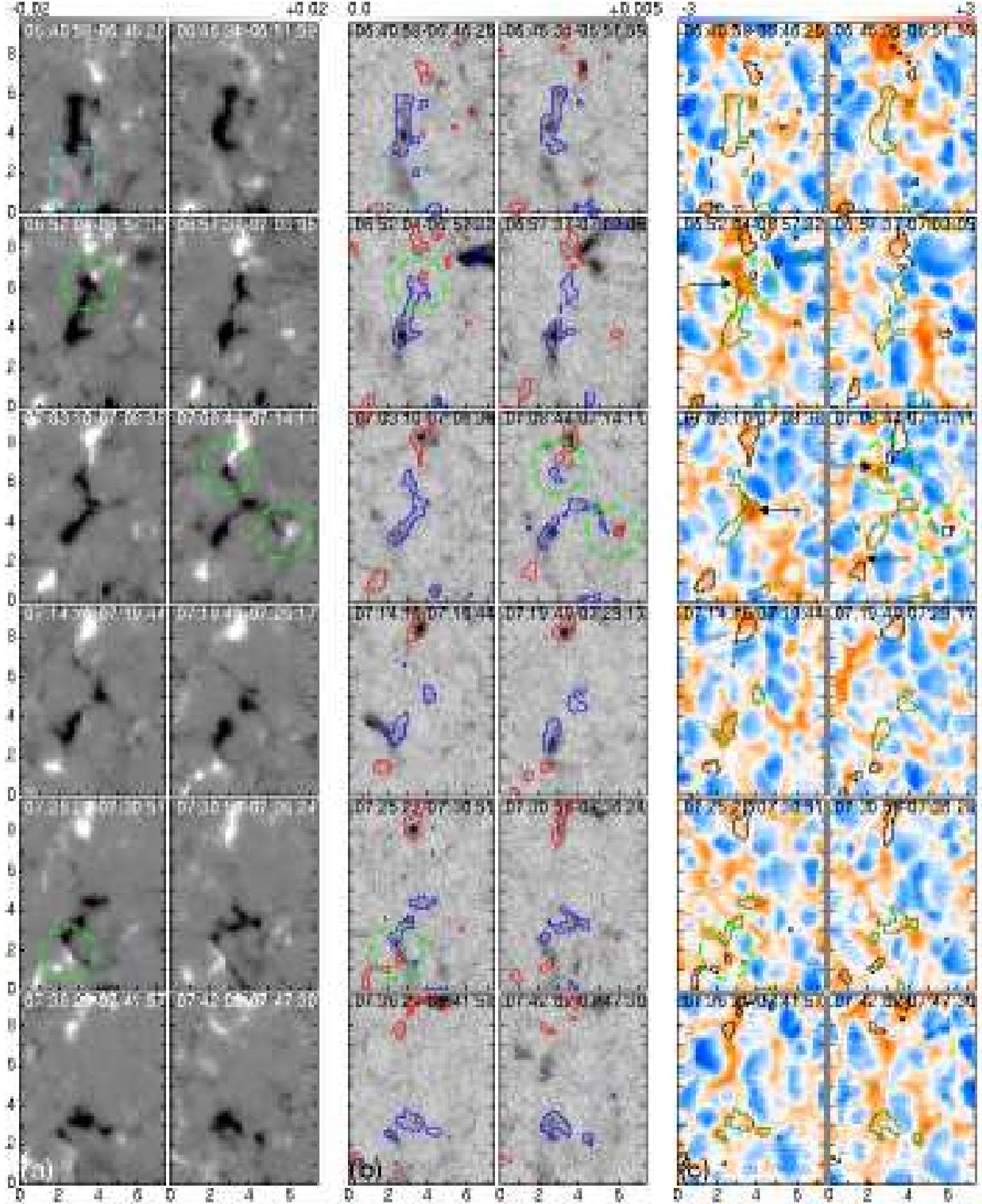}
\caption{
Time series of (a) the total circular polarization ($C_{tot}$), (b) the
 total linear polarization ($L_{tot}$), and (c) the Doppler velocity.  
The contours represent the $\pm 0.01 C_{tot}$ levels.
The dashed circles and the dashed box indicate the flux cancellation
 sites and the flux emergence site, respectively.
The arrows in panel \textit{c} point to the areas with strong red shifts.
}
\label{fig_fov2_sp}
\end{figure}

Figure~\ref{fig_fov2_sp} shows a negative element in Region B that
evolves into three branches. 
Each of these three branches elongates toward an area with a large red
shift (the arrows in Fig.~\ref{fig_fov2_sp}).  
Magnetic flux cancellations subsequently occur at the tip of each
branch (the dashed circles in Fig.~\ref{fig_fov2_sp}).  
An increase of $L_{tot}$ signals indicating horizontal magnetic fields
is observed only in the cancellation event enclosed by the final dashed
circle. 
The other 3 cancellation events do not accompany the horizontal magnetic
fields, and their properties are similar to the flux cancellation
observed in Region A. 
In the event with the increase of $L_{tot}$, the horizontal fields are
located along the intergranular lane with the Doppler red shifts.   
The sizes of the magnetic elements begin to decrease slightly in the
$C_{tot}$ maps just before their collision.
This event is different from the other cancellation events:
the horizontal field ($L_{tot}$ signals) is already present as a result
of the flux emergence well before the event of cancellation sets
in, as identified by the dashed box at the beginning of the observation
sequence on the top of Figure~\ref{fig_fov2_sp}\textit{a}.
This flux emergence has a possibility of distinguishing this event from the
other four cancellation events. 
However, it is unclear whether these horizontal fields are the same as the
horizontal fields observed during the cancellation or not, because the
$L_{tot}$ signals in the dashed box disappear briefly before the opposite
polarity elements approach each other. 

\begin{figure}
\epsscale{0.4}
\plotone{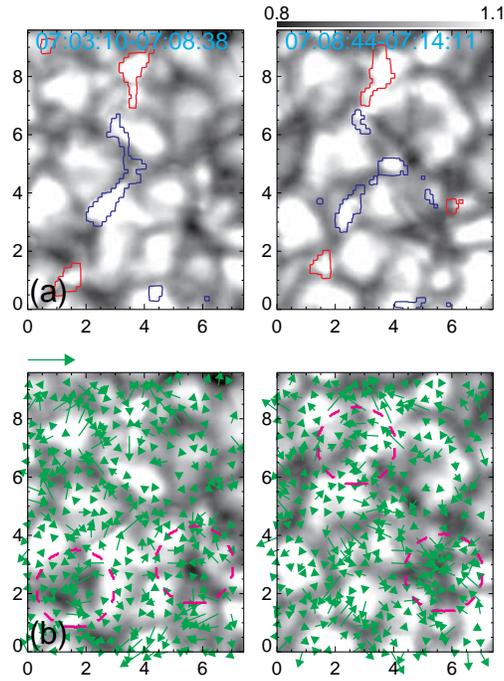}
\caption{
(a) G-band images and (b) flow maps with the local correlation tracking
 method in Region B.  
The field of view and the contours are identical to those of the third
 row in Fig.\ref{fig_fov2_sp}. 
The right-handed arrow at the top of panel \textit{b} is 1.5 km s$^{-1}$.
The dashed circles encircle the areas with strong converging flows.
The background of panel \textit{b} is identical to that of panel
 \textit{a}. 
}
\label{fig_fov2_fg}
\end{figure}

The circles in Figure~\ref{fig_fov2_fg} show that the areas with strong
converging flows mostly have a darker G-band intensity in the
intergranular lanes.
The size of these converging flow areas is larger than that of the
cancellation event in Region A, but still smaller than a travel
distance of the approaching magnetic elements.
The areas with strong converging flows correspond to the junctions of
 the intergranular lanes, and are observed to have downward Doppler motions. 
The point to emphasize is that strong converging areas do not
necessarily have cancellation events, whereas the cancellation sites
tend to be located beside or at the strong converging areas, as
suggested by our observations.

\section{DISCUSSION}
We have observed the detailed evolution of magnetic fields and velocity
fields for 5 cancellation events at the granular scales just outside the
sunspot moat region.
From the observations at high spatial resolutions, we have clearly
confirmed that opposite-polarity magnetic elements mutually cancel by
moving toward the junctions of intergranular lanes.
The intergranular junctions are characterized with darker intensities
than their nearby intergranular lanes, strong red-shifts, and surface
converging flows. 
Our new finding is that no horizontal field (no increase of $L_{tot}$) 
appears between the canceling opposite polarity in 4 of the 5 flux
cancellation sites.  
In the events without the horizontal field, the canceling
opposite-polarity magnetic elements have nearly vertical fields even
while their net magnetic flux decreases.
Here, we discuss the physical issues posed by these cancellation events,
for flux removal and flux-tube collisions as dynamical processes.

\subsection{Cancellation without Appearance of Horizontal Fields}
\subsubsection{Flux Removal from the Photosphere}
The submergence of an $\Omega$-loop is dynamically quite different from
the emergent rise of a U-loop.
Nevertheless, the photospheric convergence of loop footpoints in both
processes produces a cancellation of opposite elements with the same
magnetic signature.
In each case, the cancellation of opposite vertical fields is accompanied
with an increase followed by a decrease in the horizontal field. 
This magnetic signature is not found in our observations.
The flux cancellation process in our observations is not spatially
resolved even with the \textit{Hinode} SOT. 
In other words, flux is removed from the photosphere at sizes smaller
than the $\sim$200 km spatial resolution of the SOT.  
The observed Doppler velocities at the cancellation sites are probably
related to neither submergence nor emergence of magnetic field lines.
These velocities are just the expected downdrafts of the convective
motions at the intergranular lanes which are present even when 
flux cancellation is not occurring. 
We believe that the absence of the horizontal field in the cancellation
sites is not due to either a low cadence or an insufficient
signal-to-noise ratio (S/N).
The decrease of the canceling (colliding) magnetic elements is usually
observed in two or more frames.
There is a low possibility that the horizontal field appears only during
the period when the slit is located outside the cancellation sites.
Moreover, small-scale horizontal magnetic fields are observed outside our
cancellation sites in this data set. 
Such horizontal fields typically have an order of hecto-gauss
\citep{Orozco2007, Lites2008, Ishikawa2009}. 
Horizontal magnetic fields having sizes larger than the spatial
resolution must be detected at the cancellation site unless such
horizontal fields suddenly become one order of magnitude weaker than the
vertical fields in the colliding magnetic elements.

\subsubsection{Approach and Collision of Opposite-polarity Magnetic Elements}
Another important issue is why a magnetic polarity element would move in
such a manner as to meet with another element of the opposite polarity
on the vast solar surface. 
One possibility is the chance encounter of these elements advected by
either granular flows or supergranular flows.
The magnetic elements advected by these flows are likely to stay in
the intergranular lanes, especially the boundary of supergranular
cells. 
Flux cancellation events also prefer to occur there. 
This is consistent with our observations that magnetic flux
cancellations occur in the intergranular junctions having the strong
converging flows. 
However, one remaining issue is that motion of the approaching
opposite polarity elements is not necessarily consistent with
the surrounding surface flow patterns.
The approaching opposite-polarity elements travel a longer distance 
than the size of the strong converging areas, and do not always move 
toward the nearest, stronger converging area. 
Moreover, it is difficult to explain such motion of magnetic elements
from only the advection by supergranular flows, because these magnetic elements
move at the speed similar to usual granular flows, which is faster than 
a typical speed of supergranular flows.   
One possibility is that we have missed high-speed, systematic flows along the
intergranular lanes because of the insufficient spatial resolution in our
LCT velocity maps.
Although such flows are not yet reported, supersonic granular horizontal
flows recently detected by Doppler measurements far from the disk center
\citep{Bellot2009} may be able to produce the high-speed flows along the
intergranular lanes. 
Another possibility is that other forces (e.g., an intrinsic Lorentz force) in
addition to the force driven by the surface flows are needed to explain
the motion of the approaching magnetic elements in the photosphere.  

The subsurface advection of magnetic fields is as important as the
surface advection for thinking about the observed motion of photospheric
magnetic elements.   
In particular, if a subsurface field connects the opposite-polarity
magnetic elements as a U-loop, our observations suggest the following
interesting possibility.  
A cooler material sinking to the bottom of the U-loop would not only
prevent the loop from rising, but may even force the loop to sink with
the converging and downward convective flows, as sketched in 
Figure~\ref{fig_model}\textit{a}. 
The cooler material can drag the subsurface field lines into a deeper
layer via a downward convective flow because the plasma-$\beta$
is higher than unity below the photosphere, except for the strong magnetic 
fields associated with sunspots.
This is similar to the idea of downward flux pumping by the
turbulent granular convection around sunspots \citep{Thomas2002}.
As a result of the forced submergence of the subsurface U-loop connecting
the opposite polarity elements, these magnetic elements are driven by
their Lorentz force to move toward the area with the large downward
motion and then collide with each other (the middle panel of 
Fig.~\ref{fig_model}\textit{a}).
More detailed behaviors of flux tubes with the submergence of the
subsurface U-loop can be found in Appendix A.
Such a process is effective not only for the collision of the opposite
polarities but also for the merging of the same polarities that are 
observed before the cancellation. 
Magnetic reconnection just above or below the solar surface would be
needed for the disappearance of the colliding opposite polarity magnetic
elements from the photosphere (the right panel of
Fig.~\ref{fig_model}\textit{a}), because the bottom of the subsurface
field lines dragged by the downdrafts of cooler materials 
does not easily emerge into the photosphere.  
The reconnection site is still unknown from our observations, but
the unresolved fine-scale flux removal process at such a site suggests
magnetic reconnection, if any, is close to the photospheric surface. 
We of course have no observational evidence that a subsurface U-loop
connects the opposite-polarity elements. 
Nevertheless, the events observed in this study can be explained
by such a subsurface process.

\subsection{Cancellation with Appearance of Horizontal Fields}
The horizontal magnetic fields between the canceling opposite
polarity elements have been observed along the intergranular lanes
characterized with Doppler red shifts. 
The horizontal fields with the red shifts at the cancellation sites are
already reported by \citet{Chae2004, Cheung2008, Iida2010}. 
This cancellation is most probably a result of the submergence
(retraction) of low-lying photospheric field lines along 
the intergranular lanes (Fig.~\ref{fig_model}\textit{b}).
Such low-lying photospheric field lines probably are formed by 
magnetic reconnection in the photosphere (or the bottom of the
chromosphere) or by a failure of the emergence into the upper
atmosphere.  
Recent three-dimensional (3D) radiative MHD simulations show the
granular-scale flux cancellation due to a retraction of 
the $\Omega$-loop within an emerging flux region \citep{Cheung2008}. 
In the previous studies, the flux cancellations with the horizontal fields
are reported in the moat region \citep{Chae2004},  within
the emerging flux region \citep{Cheung2008}, or around the center of
complicated active regions \citep{Kubo2007}.
These regions basically contain many horizontal fields even if the
horizontal field is not observed just before the cancellation.
In our cancellation event with the horizontal field, small flux emergence
accompanied by the appearance of horizontal fields is also observed well
before the start of the event.
The observing products during the cancellation may depend on the
magnetic field configuration already formed before the approaching and
canceling process. 
On the other hand, the opposite polarity elements approach the region
with the converging flows, the darker intensity, and the large red
shifts as in the case of cancellations without the observation of the
horizontal fields. 
Therefore, the converging and downward motions are also important to
bring one magnetic polarity element to another polarity element in this
event.

\section{CONCLUSION}
We have presented an observational study of flux cancellation events on
the photosphere that are sufficiently resolved by the \textit{Hinode}
SOT to show that they are characterized by the absence of a horizontal field
during the cancellation process.
These events are interesting because in the usual idea of the
submergence of a low-lying $\Omega$-loop or the buoyant rise of a
U-loop, the appearance of a horizontal field is the observational
signature of the loop top (or bottom) passing across the photosphere.
Such flux cancellations appear to be more common than the
cancellation with the appearance of the horizontal fields. 
Although the nature of the flux removal process at the cancellation site
is an open question for cancellations without the appearance of the
horizontal fields, our study shows that it takes place in local areas at
scales less than the $\sim$200km resolution of the SOT and close to the
solar surface.  
The distinction between the cancellations with and without the appearance
of the horizontal fields might arise from whether the canceling opposite
polarity elements have emerged into the photosphere as a pair or not. 
However, we have investigated only 5 events that have the size less than
a few arcseconds just outside the moat region of a sunspot.
We need to investigate what really determines the appearance of the
horizontal fields between the canceling magnetic elements by using more events.
In particular, the origin of the canceling magnetic elements and their
surrounding magnetic field configuration (formation of pre-existing
horizontal fields) may be important.

We have confirmed the converging and downward convective flows are
essential for the approaching and canceling process of the
opposite-polarity magnetic elements.
The flux cancellations are observed at the intergranular junction
characterized by the strong converging and downward flows. 
However, the approaching opposite-polarity elements seem not to always
follow the surrounding surface flow patterns, at least in our observations. 
We are accustomed to thinking of the connectivities of the surface
magnetic fields that we can observe in the solar atmosphere.
Our observational study suggests that it is important to also consider the
connectivities of the surface fields that occur below the photosphere.
Lanes of downdrafts of cool fluids entraining magnetic fluxes that thread
across these lanes below the photosphere must be a common occurrence.
Such a process is a simple explanation for a pair of opposite-polarity
elements to appear to seek each other on the photosphere.
Information of subsurface convective flows would be extremely helpful for
better understanding of the magnetic flux cancellation. 

\acknowledgments
We are grateful to T. Yokoyama, Y. Katsukawa, A. de Wijn, Y. Iida, and
Y. Fan for useful discussions and comments on this paper.  
\textit{Hinode} is a Japanese mission developed and launched by
ISAS/JAXA, with NAOJ as domestic partner and NASA and STFC (UK) as
international partners. It is operated by these agencies in cooperation
with ESA and NSC (Norway). 
The FPP project at LMSAL and HAO is supported by NASA contract NNM07AA01C.

\appendix

\section{Discrete Magnetostatic Flux Tubes}

\begin{figure}
\epsscale{1.0}
\plotone{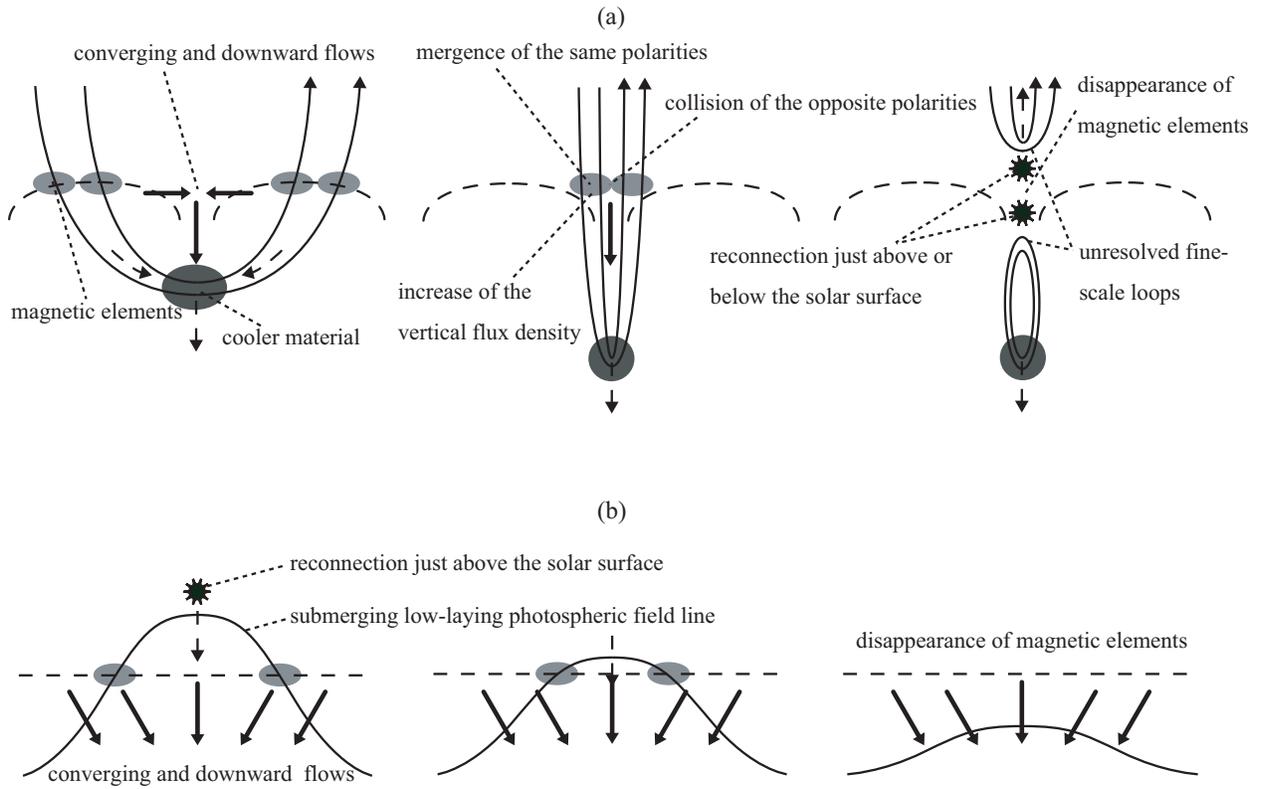}
\caption{
Schematic illustration for a flux cancellation without (panel
 \textit{a}) and with (panel \textit{b}) the observation of horizontal
 fields. 
}
\label{fig_model}
\end{figure}

The dynamical process sketched in Figure~\ref{fig_model} is due to an
interplay among pressure, gravitational and Lorentz forces in the
neighborhood of a subphotospheric convective downflow. 
A proper study of this 3D time-dependent process requires numerical MHD
simulation. 
On the other hand, some physical insight into that interplay can be seen
in 3D static solutions of discrete magnetic flux tubes in a stratified
atmosphere. 
Generally, such static solutions also require numerical computation but
the family of analytical solutions taken from Low (1982) serves our
purpose here. 
 
Consider the magnetostatic equilibrium equations:
\begin{eqnarray}
\label{mag_stat}
{1 \over 4 \pi} \left( \nabla \times {\bf B} \right) \times {\bf B} - 
\nabla p -\rho g {\hat z} = 0 , \\
\label{solenoidal}
\nabla \cdot {\bf B} = 0 , \\
\label{isothermal}
p = \rho {k T_0 \over m} ,
\end{eqnarray}
\noindent
describing the force balance for a magnetic field ${\bf B}$ in an
isothermal atmosphere at temperature $T_0$ and stratified by a uniform
gravity with acceleration $g$ the $-z$-direction, using Cartesian
coordinates. 
We use the ideal gas law (\ref{isothermal}) where $k$ and $m$ denote the
Boltzmann constant and the mean molecular weight of the plasma.
 
A 3D particular solution of these equations can be constructed for a
magnetic field of the form 
\begin{eqnarray}
\label{field}
{\bf B} & = & F(\phi, x) \nabla \phi \times \nabla x , \nonumber \\
	          & = & F(\phi, x) \left(0, {\partial \phi \over \partial z} ,-{\partial \phi \over \partial y} \right) ,
\end{eqnarray}
\noindent 
where $F(\phi,x)$ is an arbitrary function of two variables and
\begin{equation}
\label{phi}
\phi = {2 \over k_0}exp(-{1 \over 2}k_0 z) \sin \left( {1 \over 2}k_0 y \right) ,
\end{equation}
\noindent
for some constant $k_0$.  
Direct substitution of this field into the magnetostatic equations shows
that the equilibrium pressure must take the form 
\begin{equation}
\label{pressure}
p = \left( P_0 - {1 \over 8 \pi} F^2(\phi, x) \right) exp(- k_0 z) ,	         
\end{equation}
\noindent  
introducing a constant $P_0$ and identifying $k_0^{-1}=kT_0/mg$ as the
hydrostatic scale height which is of the order of $300 ~km$ at the
photosphere at a temperature of about $6000K$.

\begin{figure}
\epsscale{1.0}
\plotone{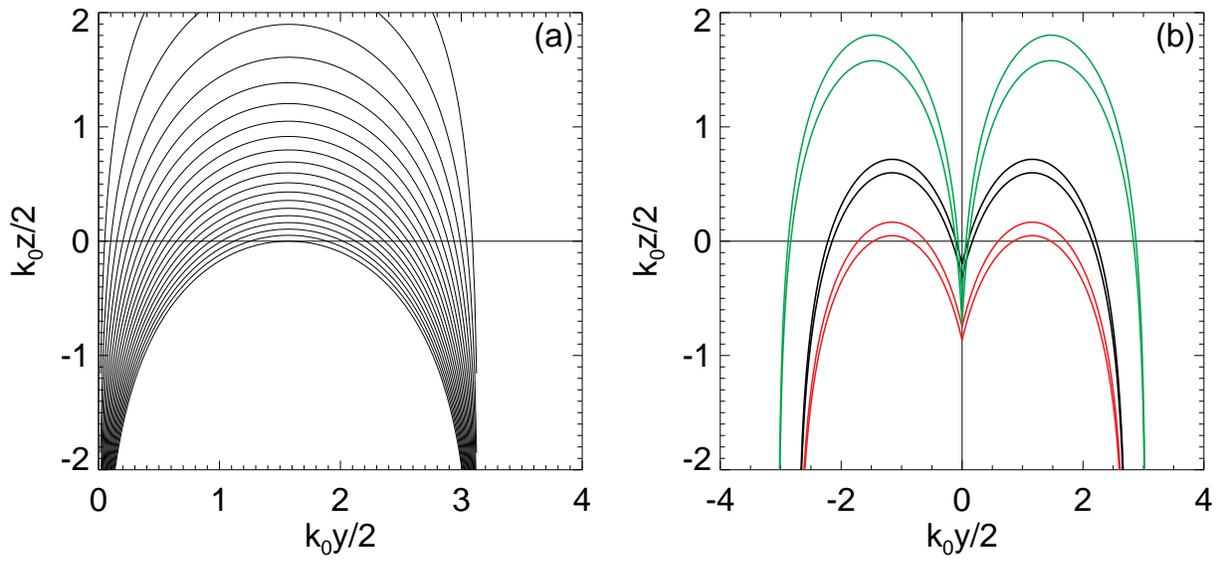}
\caption{
(a) Contours of constant $\phi(y,z)$, given by Eq.(\ref{phi}), on a plane of
 constant $x$. 
(b) Three states of the M-shaped flux tube that is composed of two
 $\Omega$-shaped flux tubes joined at a kink. 
}
\label{fig_appendix}
\end{figure}

This family of solutions is geometrically quite simple although it is
three-dimensionally varying. 
Figure~\ref{fig_appendix}\textit{a} shows the contours of constant 
$\phi(y, z)$ on a plane of constant $x$. 
The lines of force  (LOFs) are all geometrically the same on each
constant-$x$ plane but the field varies with all three Cartesian
coordinates through its amplitude function $F(\phi, x)$ which is an
arbitrary function to be prescribed.
Take any explicit function $\sigma(\phi, x)$.  
Then, setting $\sigma = \sigma_0$, a constant, generates a magnetic flux
surface in 3D space. 
To construct a flux tube of a finite cross section, a suitable
functional form of $\sigma(\phi, x)$ set to a constant $\sigma_0$
describes a flux-tube boundary. 
Then, prescribing $F(\phi, x) \ne 0$ inside the tube and $F(\phi, x)
\equiv 0$ in the rest of the atmosphere completes the construction. 
The atmospheric pressure distribution is then given by Equation
(\ref{pressure}) which can also be expressed in the form 
\begin{equation}
\label{total_P}
p + {B^2 \over 8 \pi} = P_0 \exp(- k_0 z) .
\end{equation}
\noindent
External to the flux tube, ${\bf B} = 0$ so that $p = P_0 \exp(- k_0
z)$, the isothermal pressure of the field-free part of the atmosphere.
Internal to the flux tube, the atmospheric pressure is reduced and
compensated by the magnetic pressure so that the total pressure is
stratified in the same manner as the external atmospheric pressure.
This solution is mathematically analogous to the solution describing the
equilibrium between a field ${\bf B} = B(x, y){\hat z}$ and fluid
pressure in the absence of gravity.
Equilibrium in this case is satisfied by requiring the total pressure $P
= p + B^2/8\pi$ to be uniform in space. 
We have complete freedom to prescribe the field distribution and use
this requirement to obtain the associated equilibrium pressure. 

The static balance of forces may be viewed by writing Equation
(\ref{mag_stat}) in the form 
\begin{equation}
\label{mag_stat2}
{1 \over 4 \pi} \left({\bf B} \cdot \nabla \right) {\bf B} - \nabla \left( p + B^2/8\pi \right) -\rho g {\hat z} = 0 ,
\end{equation}
\noindent   
describing the balance among the forces of magnetic tension, total
pressure, and gravity, represented by the three terms, respectively.
Since the total pressure is vertically stratified everywhere, it follows
that this class of equilibrium magnetic fields is characterized by a
tension force that is vertically oriented everywhere \citep{Low1984}.
Where the atmospheric plasma is threaded with a magnetic field, its
pressure and density are reduced, so that this portion of the atmosphere
is buoyant \citep{Parker1979}. 
The buoyancy force is just the net sum of the second and third terms in
Equation (\ref{mag_stat2}), and it is balanced by the remaining term
which is the magnetic tension force.
These solutions include linearly stable equilibrium states \citep{Low1982}. 

To address the interplay among these magnetostatic forces, let us
construct three equilibrium states, each corresponding to two
$\Omega$-shaped flux tubes that are joined below the surface $z = 0$
idealized to be the photosphere. 
These three states are displayed in Figure~\ref{fig_appendix}\textit{b}
shown as vertical sections of flux tubes boundary cut by a constant-$x$
plane. 
For each flux tube, its boundaries are drawn in different colors to
identify them. 
Each pair of boundaries is the intersection of the flux-tube surface
$\sigma = \sigma_0$ with the constant-$x$ plane.

Consider the M-shaped flux tube drawn in black, with boundaries $\phi
(y, z) = 0.4, 0.45$. 
This flux tube is made up of two identically shaped tubes that join
together below the origin where it is kinked. 
This flux tube intersects the photosphere $z = 0$ at four places, giving
two pairs of opposite-polarity magnetic elements on either sides of the
origin.
Since these elements are threaded by the same flux tube, the two
elements on either side of the origin are of opposite polarities. 
The location of these two elements is controlled by the anchoring of the
two parts of the flux tube at the kink below the origin.  The infinitely
long far arms are free to equilibrate with their surrounding fluid.
The flux tube is everywhere in equilibrium except at the kink where an
upward Lorentz force is assumed to be balanced by some force
representing the effect, for example, of the ram pressure in a
downdraft. 
The kinked part of the flux tube has a vertical thickness of about $0.05
k_0$.    

Suppose we forcefully move this kinked part of the flux tube from its
original depth at about $0.1 k_0$ below $z = 0$ to about $0.35 k_0$,
keeping the same flux distribution inside the tube and the same vertical
thickness at the kink. 
The new equilibrium of the flux tube is shown in red. 
The entire flux tube has sunk with that downward displacement of the
kinked part. 
The opposite-polarity elements on the two sides of the origin drift
apart to seek their opposite elements further away from the origin. 
This behavior can be understood as follows.  
The flux tube in this case does not extend very high into the
atmosphere. 
The tube footpoint separations are of the order of a hydrostatic scale
height, and the curvature at the tops of the tube is strong enough to
produce a tension force comparable to the pressure and buoyancy forces
at the temperature of about $6000K$ for the photosphere. 
When the kinked part of the flux tube is submerged further, an immediate
effect is to increase that magnetic curvature. 
This enhances the magnetic tension force over the buoyancy force,
leading to the submerging displacement for the whole flux tube.    

The flux tube in green shows a different behavior as the result of the
buoyancy force. In this case, the kinked part of the flux tube in 
black is submerged with two effects. 
The submergence elongates the vertical thickness of the kinked part of
the flux tube while the cross section of the tube becomes more narrow.
This elongation may be produced by a downward displacement that is
larger at the bottom than at the top of the kinked part of the tube.
The three-dimensionality of our magnetostatic solution is essential. 
If the field does not vary in $x$, we would have a flux layer rather
than a flux tube. 
The entire atmosphere lying above it is trapped and cannot fall through
without breaking the layer with variation in $x$. 
In our case, we have a true flux tube with a finite cross-sectional area.
This means that the atmosphere external to the flux tube can yield and
flow around a rising flux tube, for example.  
Moreover, the narrowing of the cross section of the tube enhances the
magnetic pressure in the tube, which naturally produces a siphon flow
along the flux tube to drain fluid down the far arms of the tube
\citep{Pikelner1971, Thomas1988}. 
The reduced weight of the tops of the tube becomes more buoyant and rise
to a greater height. 
This brings about equilibrium with the external atmospheric pressure and
a final balance between buoyancy and magnetic tension forces. 
The dynamical process builds up stress that pushes the two far arms away
from the origin while at the same time brings the two opposite-polarity
elements on either sides of the origin closer together, in a manner
similar to the \textit{Hinode} observations reported in this paper. 
In our \textit{Hinode} observations the canceling opposite-polarity elements
have fields that return to the photosphere at distances much larger than
10 hydrostatic scale heights from the canceling sites. 
We expect the buoyancy force to dominate in the atmospheric portion of
the field, keeping the tops of these fields up in the chromosphere and
above.  

This simple theoretical analysis illustrates the interplay among the
magnetic and hydrodynamic forces in the submergence of a field
connection that happens to be located right at a subphotospheric
convective downdraft. 
We of course need 3D time-dependent MHD simulations to examine the
dynamics of this process to get a complete physical picture.

\clearpage

\begin{thebibliography}{}
\bibitem[Bellot Rubio(2009)]{Bellot2009} Bellot Rubio, L.~R.\ 2009, \apj, 700, 284 
\bibitem[Bellot Rubio \& Beck(2005)]{Bellot2005} Bellot Rubio, L.~R., \& Beck, C.\ 2005, \apjl, 626, L125 


\bibitem[Chae et al.(2004)]{Chae2004} Chae, J., Moon, Y., \& Pevtsov, A.~A.\ 2004, \apjl, 602, L65 
\bibitem[Cheung et al.(2008)]{Cheung2008} Cheung, M.~C.~M., Sch{\"u}ssler, M., Tarbell, T.~D., \& Title, A.~M.\ 2008, \apj, 687, 1373 

\bibitem[de Wijn et al.(2008)]{deWijn2008} de Wijn, A.~G., Lites, 
B.~W., Berger, T.~E., Frank, Z.~A., Tarbell, T.~D., 
\& Ishikawa, R.\ 2008, \apj, 684, 1469 

\bibitem[Hagenaar(2001)]{Hagenaar2001} Hagenaar, H.~J.\ 2001, \apj, 555, 448 
\bibitem[Harvey et al.(1999)]{Harvey1999} Harvey, K.~L., Jones, H.~P., Schrijver, C.~J., \& Penn, M.~J.\ 1999, \solphys, 190, 35 

\bibitem[Iida et al.(2010)]{Iida2010} Iida, Y., Yokoyama,
			    T., \& Ichimoto, K.\ 2010, \apj, in press 
\bibitem[Ishikawa \& Tsuneta(2009)]{Ishikawa2009} Ishikawa, R., \& Tsuneta, S.\ 2009, \aap, 495, 607 

\bibitem[Jefferies et al.(1989)]{Jefferies1989} Jefferies, J., Lites, 
B.~W., \& Skumanich, A.\ 1989, \apj, 343, 920 

\bibitem[Kosugi et al.(2007)]{Kosugi2007} Kosugi, T., et al.\ 2007, \solphys, 243, 3 
\bibitem[Kubo et al.(2008)]{Kubo2008} Kubo, M., Lites, B.~W., Shimizu, T., \& Ichimoto, K.\ 2008, \apj, 686, 1447 
\bibitem[Kubo \& Shimizu(2007)]{Kubo2007} Kubo, M., \& Shimizu, T.\ 2007,
		\apj, 671, 990 

\bibitem[Lites et al.(2008)]{Lites2008} Lites, B.~W., et al.\ 2008, \apj, 672, 1237 
\bibitem[Litvinenko(1999)]{Litvinenko1999} Litvinenko, Y.~E.\ 1999, 
\apj, 515, 435 
\bibitem[Low(1982)]{Low1982} Low, B.~C.\ 1982, \apj, 263, 952 
\bibitem[Low(1984)]{Low1984} Low, B.~C.\ 1984, \apj, 277, 415 



\bibitem[Martin, Livi, \& Wang(1985)]{Martin1985} Martin, S.~F., Livi, S.~H.~B., \& Wang, J.\ 1985, Australian Journal of Physics, 38, 929 

\bibitem[Nisenson et al.(2003)]{Nisenson2003} Nisenson, P., van 
Ballegooijen, A.~A., de Wijn, A.~G., \& S{\"u}tterlin, P.\ 2003, \apj, 587, 458 

\bibitem[November \& Simon(1988)]{November1988} November, L.~J.~\& Simon, G.~W.\ 1988, \apj, 333, 427 


\bibitem[Orozco Su{\'a}rez et al.(2007)]{Orozco2007} Orozco Su{\'a}rez, D., et al.\ 2007, \pasj, 59, 837 

\bibitem[Parker(1975)]{Parker1975} Parker, E.~N.\ 1975, \apj, 201, 494 
\bibitem[Parker(1979)]{Parker1979} Parker, E.~N.\ 1979, Oxford, Clarendon Press; New York, Oxford University Press, 1979, 858 p.,  
\bibitem[Pikel'Ner(1971)]{Pikelner1971} Pikel'Ner, S.~B.\ 1971, \solphys, 17, 44 

\bibitem[Ryutova et al.(2003)]{Ryutova2003} Ryutova, M., Tarbell, T.~D., \& Shine, R.\ 2003, \solphys, 213, 231 

\bibitem[Schrijver et al.(1997)]{Schrijver1997} Schrijver, C.~J., Title, A.~M., van Ballegooijen, A.~A., Hagenaar, H.~J., \& Shine, R.~A.\ 1997, \apj, 487, 424 
\bibitem[Shimizu et al.(2008)]{Shimizu2008} Shimizu, T., et al.\ 2008, \solphys, 249, 221 
\bibitem[Stenflo(1989)]{Stenflo1989} Stenflo, J.~O.\ 1989, \aapr, 1, 3

\bibitem[Takeuchi \& Shibata(2001)]{Takeuchi2001} Takeuchi, A., \& 
Shibata, K.\ 2001, \apjl, 546, L73 
\bibitem[Thomas(1988)]{Thomas1988} Thomas, J.~H.\ 1988, \apj, 333, 407 
\bibitem[Thomas et al.(2002)]{Thomas2002} Thomas, J.~H., Weiss, 
N.~O., Tobias, S.~M., \& Brummell, N.~H.\ 2002, \nat, 420, 390 
\bibitem[Tsuneta et al.(2008)]{Tsuneta2008} Tsuneta, S., et al.\ 2008, \solphys, 249, 167 

\bibitem[Yurchyshyn \& Wang(2001)]{Yurchyshyn2001} Yurchyshyn, V.~B.~\& Wang, H.\ 2001, \solphys, 202, 309 

\bibitem[Zwaan(1987)]{Zwaan1987} Zwaan, C.\ 1987, \araa, 25, 83 

\end{thebibliography}
\end{document}